%% file: ms.tex
\documentclass{emulateapj}

\shorttitle{The magnetic field in \object{K~3-35}}
\shortauthors{G\'omez et al.}

\begin{document}


\title{The magnetic field toward the young planetary nebula \object{K~3-35}}


\author{Y. G\'omez, and  D. Tafoya\altaffilmark{1}}
\affil{Centro de Radioastronom\'\i a y Astrof\'\i sica, UNAM,
    Apartado Postal 3-72, Morelia, Michoac\'an 58089, Mexico.}
\email{y.gomez@astrosmo.unam.mx, d.tafoya@astrosmo.unam.mx}

\author{G. Anglada and L. F. Miranda}
\affil{Instituto de Astrof\'\i sica de Andaluc\'\i a, CSIC, Apdo. Correos 3004,
E-18080 Granada, Spain}
\email{guillem@iaa.es, lfm@iaa.es}

\author{J. M. Torrelles}
\affil{Instituto de Ciencias del Espacio (CSIC)-IEEC, Facultat de Fisica, Universitat 
de Barcelona, Planta 7a, Marti i Franqu\`es 1, 08028 Barcelona, Spain}
\email{torrelles@ieec.fcr.es}

\author{N. A. Patel}
\affil{Center for Astrophysics, 60 Garden Street,
Cambridge, MA 02138, USA}
\email{npatel@cfa.harvard.edu}

\and

\author{R. Franco Hern\'andez\altaffilmark{1}}
\affil{Centro de Radioastronom\'\i a y Astrof\'\i sica, UNAM,
    Apartado Postal 3-72, Morelia, Michoac\'an 58089, Mexico.}
\email{r.franco@astrosmo.unam.mx}


\altaffiltext{1}{Predoctoral Student at the Center for  Astrophysics, 60 Garden Street,
Cambridge, MA 02138, USA}


\begin{abstract}
\object{K~3-35} is a planetary nebula (PN) where H$_2$O maser 
emission has been detected, suggesting that it  
departed from the proto-PNe phase only some decades ago.
Interferometric VLA observations of the OH 18~cm transitions 
in \object{K~3-35} are presented. OH maser emission is detected in all 
four ground state lines (1612, 1665, 1667, and 1720 MHz). 
All the masers appear blueshifted with respect to the systemic velocity of the
nebula and they have different spatial and kinematic distributions.
The OH 1665 and 1720 MHz masers appear spatially coincident with 
the core of the nebula, while the OH 1612 and 1667 MHz ones exhibit a 
more extended distribution. 
We suggest that the 1665 and 1720 masers arise from a region 
close to the central star, possibly in a torus, while the 1612 and 1667 
lines originate mainly from the extended northern lobe of 
the outflow.
It is worth noting that the location and velocity of the OH 1720 MHz maser
emission are very similar to those of the H$_2$O masers 
(coinciding within 0${\rlap.}^{\prime\prime}$1 and $\sim$2 km~s$^{-1}$, respectively). 
We suggest that the pumping mechanism in the H$_2$O masers could
be produced by the same shock that is exciting the OH 1720 MHz transition.
A high degree of circular polarization ($>$50\%) was found to 
be present in some 
features of the 1612, 1665, and 1720 MHz emission.
For the 1665 MHz transition at $\sim$+18 km~s$^{-1}$ the emission with 
left and right circular polarizations (LCP and RCP) 
coincide spatially within a region of 
$\sim$0${\rlap.}^{\prime\prime}$03 in diameter.
Assuming that these RCP and LCP 1665 features come from a Zeeman pair, we 
estimate a magnetic field of $\sim$0.9 mG within 150 AU from the 1.3 cm 
continuum 
peak. This value is in agreement with a solar-type magnetic field associated 
with evolved stars.  

\end{abstract}





\section{Introduction}

\object{K~3-35} (PN\,G056.0+0.20; IRAS~19255+2123) is an extremely young PN 
where we are observing the first stages of formation of collimated 
bipolar outflows \citep{MI98}. Its
radio continuum emission at 3.6~cm exhibits an elongated, filamentary 
S-shape with a
remarkable point-symmetric structure with respect to the center \citep{aa93,
MI01, gom03}. Recently, its morphology has been 
successfully modeled with a precessing jet evolving in a dense asymptotic giant
branch (AGB) circumstellar 
medium \citep{vel07}. The extension of the
radio continuum jets at 3.6~cm is $\sim$2$^{\prime\prime}$, which is 
equivalent to 
10,000 AU \citep[assuming a distance of 5 kpc;][]{zha95}.
A short dynamical age ($<$50 years) for the jets was estimated 
from \citet{vel07}, which is comparable with that of the core.
In addition, the high ratio of molecular to ionized mass \citep{taf07} 
indicates that \object{K~3-35} departed from
the proto-PN phase only some decades ago.

\object{K~3-35} was the 
first PN known to exhibit H$_2$O maser emission \citep{MI01}.
Since then, two other PNe have been reported with H$_2$O maser emission:  
IRAS~17347$-$3139 \citep{deGre04} and IRAS~18061$-$2505 \citep{gom08}.  
The water masers toward \object{K~3-35} are located at 
the center of the nebula, along the minor axis, at a radius of $\sim$85~AU and also 
at the tips of the jet lobes \citep{MI01}. 
\citet{usc08} have analyzed the kinematics of the H$_2$O masers in 
\object{K~3-35},
identifying the presence of a rotating and expanding ring with a radius 
of $\sim$100 AU, which may be related with the collimation of the outflow.

The mechanism that generates collimated outflows in PNe and their
role in the shaping of these objects is still a matter of
debate. Magnetic fields have been suggested to play a
major role in these aspects; their existence in 
circumstellar envelopes has been 
invoked to explain jets and  bipolar morphologies in PNe 
\citep[e.g.][]{rf96, GS99, GS06, mat00, mat04, bla01, GS05, fra07}. 
\citet{GS00} have modeled morphologies of PNe with and
without magnetic fields, showing the importance of magnetic fields to
produce collimated ejections.
\citet{sok06} has argued that the magnetic fields in PNe cannot shape 
the morphology of the nebula alone without the presence of a stellar
companion that produces a spin-up mechanism in the envelope. 
The detection of magnetized disks toward evolved stars is therefore crucial
to understand the generation of collimated jets and bipolar structures
in PNe.

Magnetic fields have been detected in several proto-PNe 
\citep{ZI89, kd97, vle05, vle06, ban03, et04, szy05, her06, vle08}
but only toward a few PNe \citep{MI01, gre02, jor05}. The strength of 
the magnetic fields detected
in envelopes of evolved objects ranges from 1 G, at a radius of r$\sim$1~AU, 
to 10$^{-4}$~G, at r$\sim$1000~AU, and is of the order of kG in 
the central stars 
of PNe \citep{jor05}. Recently,
the geometry of the magnetic field has been inferred for the proto-PN W43A 
\citep{vle05} and the PNe NGC~7027, NGC~6537, and NGC 6302 \citep{gre02, 
sab07} suggesting the presence of toroidal magnetic fields.
In the particular case of \object{K~3-35}, it shows polarized 
OH maser emission in the 1665~MHz line around the central region, with a high 
level of circular polarization 
above $\simeq$50\%, suggesting the presence of a magnetic field 
\citep{MI01, gom05}.

In this work we present VLA\footnote{The Very Large Array (VLA) 
is operated by the National Radio Astronomy Observatory (NRAO), 
facility of the National Science Foundation operated under 
cooperative agreement by Associated Universities,~Inc.}
observations toward \object{K~3-35} of the
OH maser emission in its four ground-state transitions (1612, 1665, 1667 and 1720 MHz)
in order to study the spatial distribution and polarization
of the masers as well as the magnetic field in this PN.

\section{Observations}


The OH maser observations towards 
\object{K~3-35} at 1612.231, 1665.401, 1667.359 and 1720.530 MHz were 
carried out on 2002 March 31 with the VLA.
The array was in the A configuration giving an angular resolution of
$\sim$1$^{\prime\prime}$ at 18~cm. 
The 1665 and 1667 OH main lines were observed with a
total bandwidth of 195.31 kHz, divided into 256 channels of 0.763 kHz each.
The velocity resolution for these transitions 
was $\sim$0.14 km~s$^{-1}$ covering
a total velocity range of 35 km~s$^{-1}$. 
The 1612 and 1720 OH satellite lines were observed with a total
bandwidth of 781.25 kHz, divided into 256 channels of 3.05 kHz each, 
achieving a velocity resolution of $\sim$0.56 km~s$^{-1}$ yielding a total 
velocity range of $\sim$140 km~s$^{-1}$. 

The calibration, deconvolution, and imaging
of the data were carried out using the Astronomical Image Processing System
(AIPS) of the NRAO.  Both the right circular polarization (RCP) and 
the left circular polarization (LCP) were observed simultaneously using the 
normal spectral line mode centered
at a v$_{LSR}$ = +15 km~s$^{-1}$ for all transitions, except the 1667~MHz line
which was centered at +4 km~s$^{-1}$. We applied Hanning smoothing to all
four OH transitions in order to improve the signal-to-noise ratio of the data,
and to minimize the Gibbs phenomenon. The resulting $rms$ per channel is
$\sim$ 6 mJy~beam$^{-1}$ for the 1665, 1667, and 1720 MHz transitions and 
$\sim$ 30 mJy~beam$^{-1}$ for the 1612 MHz transition. The $rms$ for the 1612 MHz
transition is higher due to strong on-line flagging and a shorter integration
time.  
The absolute amplitude calibrator was 1331+305, the phase calibrator 
1925+211, and the bandpass calibrator 0319+415.  The flux densities of these 
calibrators for each frequency are summarized in Table~\ref{tbl-1}.
The RCP and LCP data were calibrated separately and later combined during
the imaging process to make Stokes I = (I$_{\rm RCP}$ + I$_{\rm LCP}$)/2 and
Stokes V = (I$_{\rm RCP}$ $-$ I$_{\rm LCP}$)/2 data sets, where 
I$_{\rm RCP}$ and
I$_{\rm LCP}$ are the intensities in the right and left circular polarizations,
respectively. 
Cleaned images for the RCP and LCP data were made using the task IMAGR 
of AIPS with the ROBUST parameter \citep{BR95} set to $0$, 
allowing to determine 
the peak position of the OH masers, at each velocity channel, 
with a relative accuracy of $\sim$ 0${\rlap.}^{\prime\prime}$05. 




\section{Results and Discussion}

\subsection{Continuum and OH Maser Emissions}

Figure~1 shows the 18~cm continuum emission toward \object{K~3-35}, 
taken from the 
line-free channels of the OH spectral data. The continuum emission has a
deconvolved angular size of   
1${\rlap.}^{\prime\prime}$8 $\times$ 0${\rlap.}^{\prime\prime}$5, 
P.A.= 16$^\circ$. This 18-cm continuum image shows a northeast-southwest 
elongation which resemble the orientation of the bipolar outflow 
\citep{aa93, MI01}. 
The total continuum flux density at 18~cm is 18 $\pm$ 1 mJy with a peak 
position at 
$\alpha$(2000)= 19$^h$ 27$^m$ 44${\rlap.}^{s}$026 $\pm$ 0${\rlap.}^{s}$001, 
$\delta$(2000)= 21$^{\circ}$ 30$^{\prime}$ 03${\rlap.}^{\prime\prime}$57 $\pm$
0${\rlap.}^{\prime\prime}$04. 

The 18~cm nominal peak position is located toward the north, about 
$\sim$0${\rlap.}^{\prime\prime}$13, of the 1.3~cm peak position reported 
by \citet{MI01}. We attribute this shift in position to the fact that
the 18~cm emission might be optically thick, while the 1.3 cm is 
optically thin, then 
the 1.3~cm peak may appear closer to the central star than the 18 cm peak. 
However, since the uncertainty in the absolute position for the 18 cm data is about 
0${\rlap.}^{\prime\prime}$1,  observations with higher absolute positional accuracy
are needed to confirm this spatial shift.

The OH 1665 and 1667 MHz main lines, as well as the OH 1612 and
1720 MHz satellite lines were all detected and imaged toward \object{K~3-35}. 
Figure~2 shows the RCP, LCP, I and V spectra made with the task ISPEC
of AIPS, for the four ground-state lines.
The detected OH maser emission lines appear with LSR velocities
$<$23 km~s$^{-1}$, which are all blueshifted
with respect to the systemic velocity of the source derived from the CO (2-1)
molecular gas \citep[$\sim$23 km~s$^{-1}$;][]{taf07}. 
In general, the OH maser spectrum in an evolved star is characterized by two
peaks which are blueshifted and redshifted, coming mainly from the front and 
back parts of the expanding envelope, respectively \citep{RE77}. 
In the presence of an ionized region, which is the case for \object{K~3-35}, 
and since the ionized gas can become optically thick at 
low frequencies, it is possible that the redshifted maser components, coming 
from the back side of the nebula, are not detected because the 
emission could be absorbed by the free-free opacity of the 
ionized gas \citep{RGG85}.
This effect may explain why in \object{K~3-35} all the OH masers
are blueshifted with respect to the systemic velocity.

In Table \ref{tbl-2} we list the velocity, position, and 
flux density of the observed features in the OH RCP and LCP 1665, 1667, 
1612 and 1720 MHz spectra (see Figure~2). We define a maser feature as all the
spectral channels in which maser emission is detected, indicated by the properties of
its peak channel. Also, we define a maser spot as one spectral channel of a maser 
feature.
The velocities and spatial distribution of the OH maser spots are
shown in Figure~3. The positions were obtained by 2D Gaussian fitting 
in each velocity channel where the OH maser emission was above 5-$\sigma$ 
level, with a typical uncertainty of 0${\rlap.}^{\prime\prime}$05 
(see Table~\ref{tbl-2}).
Only one single maser component was fitted per channel and in all cases it is 
spatially unresolved ($\leq$1$^{\prime\prime}$).

It can be noticed from Figure~3, that the masers for the different transitions
are located toward different regions of the nebula. 
In particular, the OH 1665 and 1720 masers have a compact distribution, 
around the 1.3~cm continuum emission peak,
compared with the 1612 and 1667 masers that appear in a more 
extended area (see Figure~3). On the other hand, the OH 1665 and 1667 MHz masers 
are tracing different kinematic components, with the 1667 
maser spots having velocities lower than 10 km~s$^{-1}$. 
In what follows, we present an analysis of the spatial distribution, kinematics and
polarization for each individual OH maser transition.

\medskip

\subsubsection {1665 MHz}

\citet{ENG85} did not detect the OH 1665 MHz transition toward
\object{K~3-35} with a 3-$\sigma$ sensitivity limit of 0.6 Jy.
\citet{MI01} detected and imaged for
the first time the 1665 MHz transition toward \object{K~3-35} using the VLA,
reporting the presence of emission in a velocity range from
+14 to +20 km~s$^{-1}$. In this work (Figure~2 and Table~\ref{tbl-2}) we confirm
the presence of OH 1665 MHz maser emission, which covers the same LSR velocity 
range as reported by \citet{MI01}. The velocity, position, and flux density of the 
OH 1665 MHz RCP and LCP spectral features are listed on Table~\ref{tbl-2}.  
The location of the OH 1665 maser spots (with signal-to-noise ratios $>5$), appear 
slightly shifted to the south of the 18~cm radio 
continuum peak (see Figure~3). Almost all these maser spots fall within 
0${\rlap.}^{\prime\prime}$1 of the position of the 1.3 cm radio continuum peak, 
which we think is closer to the central star. Also, almost all the OH 1665 MHz 
features exhibit a high degree of circular polarization above 
50$\%$ and, in the particular case
of the spectral feature in the velocity range from +17.5 to +18.0 km~s$^{-1}$, the
RCP and LCP emission is spatially coincident, suggesting the presence of
a Zeeman pair.

\subsubsection {1667 MHz}

Two weak 1667 MHz spectral components toward \object{K~3-35} were first reported
by \citet{ENG85} at LSR velocities of $-$4 and +7.8 km~s$^{-1}$, and 
intensities of 0.3 and 0.4 Jy, respectively.
\citet{TLH91} did not detect the OH 1667 MHz masers
with an upper limit of 0.2 Jy. \citet{MI01} reported the detection 
of 1667 MHz maser emission in two velocity ranges, 
from $-$2 to $-$3 km~s$^{-1}$ and from +7 to +9 km~s$^{-1}$,  with 
peak flux densities lower than 0.1 Jy. In this work we report the detection
of two 1667 MHz spectral features at $-$3.5 and +8.0 km~s$^{-1}$ 
(see Figures 2, 3, and Table~\ref{tbl-2}), that are spatially
separated in two groups by about 0${\rlap.}^{\prime\prime}$1, 
toward the northeast and the southwest, 
respectively (see blue and green features in Figure~3).
There is no significant circular polarization in the OH 1667  MHz masers 
($\leq$16$\%$; see Table~\ref{tbl-2}) in agreement with the results of 
\citet{MI01}.
The two features are blueshifted with respect to the systemic velocity
of the PN ($\sim$ +23 km~s$^{-1}$), suggesting that they are arising
from the blueshifted lobe of the outflow. In any case, it is clear
from Figure~3, that the 1667 maser emission is tracing a different
region that the 1665 one.

\subsubsection {1612 MHz}

The OH 1612 MHz line was first detected toward
\object{K~3-35} by \citet{ENG85} using the
Effelsberg 100~m telescope. Three velocity features (at $-$4, +9
and +20 km~s$^{-1}$) were reported, with the strongest feature 
($\sim$ 4 Jy) at +9 km~s$^{-1}$.
\citet{TLH91}, using the Nan\c{c}ay radio telescope,
observed two 1612 MHz velocity features at $\sim$ +9 and
+20 km~s$^{-1}$.
High spectral resolution 1612 MHz observations, carried out 
also with the Nan\c{c}ay radio telescope by \citet{szy04},
showed the presence of four velocity features (at $-$2, +9, 
+18 and +21 km~s$^{-1}$). Interferometric
VLA observations revealed the coincidence in position 
(within $\sim$0${\rlap.}^{\prime\prime}$5) 
of the 1612 MHz maser emission and the continuum
emission peak, supporting the association of the OH masers with 
\object{K~3-35} \citep{aa93}.
In this work we confirm that the OH 1612 maser emission  is associated with 
\object{K~3-35} (Figure~3). Four OH 1612 maser spectral 
features were detected at $-$2.0, +8.8, +18.4 and +21.2 km~s$^{-1}$
(Figure~2 and Table~\ref{tbl-2}), in agreement with those reported 
by \citet{szy04}.
We note that the OH 1612 maser spots
are displaced in position by about 0${\rlap.}^{\prime\prime}$1 to 
the south of the 1.3~cm continuum peak  and that they show an 
extended distribution in space (see Figure~3). 

In Figure~2 we identify four main groups of 1612 MHz maser features. 
Two of them coincide in velocity with those of the 1667 MHz transition 
at $\sim -$3.5 and +8 km~s$^{-1}$, although only the group at 
+8 km~s$^{-1}$ coincides spatially
with that of the 1667 MHz transition (see green color maser spots in Figure 3),
suggesting that both the OH 1612 and 1667 velocity transitions are arising in 
the same region, (i.e. in the northern blueshifted lobe of the outflow)
\citep{MI00}.
The third and fourth groups of OH 1612 MHz masers at $\sim$ +18 and +21 
km~s$^{-1}$ appear in the
same velocity range as the 1665 transition, and it is more likely that they
are arising from an inner region of the nebula, close to the central star.   
All the features show circular polarization, with the highest polarized 
feature at +21 km~s$^{-1}$ (orange red color spots in Figure~3). 

\subsubsection {1720 MHz}

The OH 1720 MHz maser emission has been detected previously toward \object{K~3-35}
by \citet{TLH91} when the classification of this object was still
unclear. Now it is known that \object{K~3-35} is a young PN 
\citep{MI01} and so far it is the only one that exhibits
the OH 1720 MHz maser transition.  
\citet{TLH91} detected a single peak spectral component toward
\object{K~3-35} centered at +22.1 km~s$^{-1}$ with a total flux density of 3 Jy.
Our 1720 MHz detection (see Figure~2) shows a narrow velocity feature
centered at an LSR velocity of +21.4 km~s$^{-1}$ (0.6 Jy),
coincident in position with the radio continuum peak at 1.3~cm (red color maser spots
in Figure~3), suggesting that these masers originate very close to the
central star.

The OH 1720 MHz maser emission is commonly 
associated with shocked molecular regions in
star forming regions and supernovae remnants \citep{FGS94}.
Recently, the 1720 MHz transition has been detected
toward a few post-AGB stars \citep{sch01, dea04}.
\citet{eli76} has proposed that the only way to produce strong 
OH 1720 MHz maser emission is by means of collisional pumping
under particular physical conditions 
($T_{\rm k} \leq$ 200~K; n$_{\rm H_2} \simeq$10$^5$ cm$^{-3}$). 
\citet{sch01} detected OH 1720 MHz maser emission toward
the post-AGB star IRAS~18043$-$2116, and explained this emission as due to  
the passage of a C-type shock through the remnant AGB envelope caused by
the fastwind. According to the scenario proposed by these authors, H$_2$O 
molecules would be formed behind the shock. Subsequently, the H$_2$O is 
photodissociated enhancing the OH abundance. The post-shock physical
conditions are favorable for pumping the OH 1720 MHz maser transition.
Even though these authors predicted that H$_2$O masers would not be present,
\citet{dea04} detected H$_2$O maser emission toward IRAS~18043$-$2116. 
These results
suggest that the physical conditions in this source can be adequate to
excite both the OH~1720 MHz and the H$_2$O maser emissions. 
 
For \object{K~3-35} we note that the velocity and location of the OH 1720 MHz 
masers are very similar to those of the H$_2$O masers within 
0${\rlap.}^{\prime\prime}$1, which correspond to the absolute positional 
accuracy for the OH masers. This suggest that
the pumping mechanism of the H$_2$O masers could be produced by the same 
shock that is exciting the OH 1720 MHz transition. 
The OH 1720 MHz maser emission has a high degree of circular 
polarization, $\sim$50$\%$ 
(Table~\ref{tbl-2}); however, although the RCP and LCP maser features 
coincide in position (within 0${\rlap.}^{\prime\prime}$04), they do not 
show the typical ``S'' profile, indicative of a Zeeman 
pattern in the Stokes parameter V.

\medskip

\subsection{Magnetic field in K~3-35}

Polarization studies have been carried out
toward proto-PNe using OH maser transitions 
\citep[e.g.][]{ban03, szy04},
H$_2$O masers \citep{vle05, vle08}
and SiO masers \citep{her06}, showing 
for some objects magnetic field components which are orthogonal to the major
axis of the nebulae. 
Several PNe have been reported to harbor OH
maser emission \citep{pay88, ZI89, bow89, TLH96}. However, the relatively
weakness of the OH maser 
features difficults the detection of polarization using OH maser techniques.
In the case of \object{K~3-35}, the OH maser emission is strong enough
to allow us a detailed study of the polarization of this emission. 

The existence of strong circular polarization in some features of the 
OH 1612, 1665 and 1720 MHz maser transitions suggests
the presence of a magnetic field in \object{K~3-35}.
In general, the field leads to a splitting of the 
magnetic substates of the OH molecule, which can be detected as a Zeeman pair 
\citep{dav74, cru83}. 
Assuming that the source is permeated with the same strength of 
field then, if a Zeeman effect is present, the LCP and RCP 
components have to be 
emitted from the same region of the OH source. 

Figure~4 shows a close up of the RCP and LCP 1665 MHz maser spots plotted 
in Figure~3 (upper left panel) but with maser spots that have flux 
densities above 15$\sigma$. 
These maser spots are mainly distributed in three spatial regions with
different characteristic velocities: 
one close to 
the 1.3~cm continuum emission peak (marked with the upper dash-line square 
in Figure~4) and 
the other two located toward the southeast (left dash-line square) and 
southwest (right dash-line circle) 
of the 1.3~cm continuum peak. 
The region associated with the 1.3~cm continuum peak corresponds to the main
spectral feature, at $\sim$+17.1 km~s$^{-1}$ of the RCP spectrum shown in 
Figure~2 (upper left panel). 
The two regions to the southeast and southwest 
are separated in position by about 0${\rlap.}^{\prime\prime}$04,
and correspond to the two spectral features at $\sim$+15.0 and +18.6 km~s$^{-1}$, 
respectively, of the LCP spectrum shown in Figure~2 (upper left panel). 
The relative position accuracy between the LCP 1665 masers is
less than 0${\rlap.}^{\prime\prime}$006, and the resulting velocity resolution
(after Hanning smoothing) is $\sim$0.3 km~s$^{-1}$, then we consider that this 
positional and velocity shift is real. 
The shift between these two groups of OH 1665 features  
may be simply due to inhomogenities of the gas at different velocities or they 
can reveal a true kinematic effect of a single structure.  If we try to 
extrapolate the kinematics of the ring traced by the H$_2$O masers
\citep{usc08}, we note that the sense of rotation for the H$_2$O ring is
not in agreement with the velocity gradient of the two OH 1665 MHz groups
(see bottom left panel of Figure~4). 
We suggest that only the group of masers around +18.6 km~s$^{-1}$ could be associated
with dense gas close to the central star, likely in an equatorial 
torus, but the OH
1665 masers at $\sim$+15 km~s$^{-1}$, would be associated with the blueshifted lobe of
the outflow as in the case of the 1667 and 1612 MHz masers.

We also note that the positions of the 1665 RCP and LCP maser spots with velocities
$\sim$ +17.5 to +18.0 km~s$^{-1}$, overlap each other (see 
dash-line circle in upper-left panel of Figure~4). 
In this sense, the OH 1665 transition has a velocity component
that seems to follow the physical requirements for a Zeeman pattern
(they exhibit the typical ``S'' shape in the Stokes V spectrum shown in
Figure~2). Assuming that these masers are arising from the same 
spatial region (within 0${\rlap.}^{\prime\prime}$03) and that we have 
at least one Zeeman pair in the velocity range between +17.5 to  
+18.0 km~s$^{-1}$,  we can estimate the magnitude of the magnetic field 
along the line-of-sight (B$_{LOS}$). 

It is known that when the Zeeman splitting is small compared to the 
line width, the Stokes V spectrum is given by
V = dI/d$\nu$~$b$~B$_{LOS}$ Hz~$\mu$G$^{-1}$ \citep{hei93, RQC08}, 
where dI/d$\nu$ is the frequency derivative of the Stokes I spectrum
and $b$ is the splitting coefficient \citep[$b$=3.270 for the OH 1665 MHz 
transition;][]{dav74}.
In terms of the velocity, the previous expression will be 
V = 3.27~$c$/$\nu_0$~dI/dv~B$_{LOS}$ Hz~$\mu$G$^{-1}$, where $c$=3$\times$10$^5$ 
km~s$^{-1}$ and $\nu_0$=1665$\times$10$^6$~Hz. 
By comparing the Stokes V spectrum with the derivative of the Stokes I
spectrum, we obtain V=0.55~km~s$^{-1}$~dI/dv. From this scaling,  
we estimate B$_{LOS} \simeq$ 0.9 $\pm$ 0.1 mG, at a 
distance, from the 1.3 cm continuum peak, of 
$\sim$0${\rlap.}^{\prime\prime}$03 ($\sim$ 150 AU).

A comparison between the magnetic field estimates toward evolved 
stars as function of 
distance from the star, assuming  a $B\propto$r$^\alpha$ dependence, 
has been made by \citet{vle05}, plotting  
two models: a solar-type magnetic field ($\alpha$=$-$2), and a dipole 
magnetic field ($\alpha$=$-$3). Plotting the strength of the magnetic 
field of \object{K~3-35}
in figure~6 of \citet{vle05}, shows that
it is in agreement with a solar-type model.

\medskip

\section{Conclusions}

Using the VLA we detected and imaged the 1612, 1665, 1667 and 1720 MHz OH masers 
toward \object{K~3-35}.
The velocity and position distribution of the masers suggest that they are
arising from different regions in the nebula. The 1665 and 1720 MHz maser 
features
have a compact distribution, near the 1.3~cm continuum emission peak, and it
is possible that some features of the 1665 MHz transition are originating in an
equatorial disk. The 1612 and 1667 MHz maser features are spread in a 
more extended 
region in two main velocity groups. We suggest that the velocity component at
$\sim$ +8.0 km~s$^{-1}$ in the 1612 and 1667 MHz spectra could be tracing 
the same region near
the central core in the blueshifted lobe of the outflow. 
As far as we know, \object{K~3-35} is the only PN that exhibits simultaneously
OH 1720 MHz  and H$_2$O maser emission. 
We note that the velocity and location of the OH 1720 MHz
masers are very similar to those of the H$_2$O masers. This suggests that
the pumping mechanism of the H$_2$O masers could be produced by the same
shock that is exciting the OH 1720 MHz transition.
The 1612, 1665, and 1720 MHz spectra show some velocity features with 
a high degree of circular polarization ($>$50\%). In the particular
case of the velocity feature at $\sim$ +18~km~s$^{-1}$ in the OH 1665 MHz 
transition, 
it shows the typical ``S'' profile for the Stokes V, and the
LCP and RCP features  are arising 
from the same spatial position (within $\sim$0${\rlap.}^{\prime\prime}$03 
in diameter).
This result suggests that a Zeeman pair could be present in \object{K~3-35} and a 
magnetic field in the line-of-sight of
$\sim$0.9 mG, at a radius of 150 AU, is estimated. This value 
is in agreement with a solar-type ($\alpha$=$-$2, B$\propto$r$^\alpha$) 
magnetic field associated with evolved stars. 
\medskip

\acknowledgments

YG, DT and RFH acknowledge support from DGAPA, UNAM, and CONACYT, Mexico.
GA and JMT acknowledge support from MEC (Spain) AYA 2005-08523-C03 and
MICINN (Spain) AYA 2008-06189-C03 grants (co-funded with FEDER founds). 
LFM is supported partially by grant AYA2005-01495 of the Spanish MEC and
grant AYA2008-01934 of the Spanish MICINN (both co-funded by FEDER funds).
GA, LFM, and JMT acknowledge partial support from Junta de Andaluc\'\i a (Spain).

\clearpage



\begin{figure}
\begin{center}
\vspace{3cm}
\includegraphics[angle=0,scale=0.8]{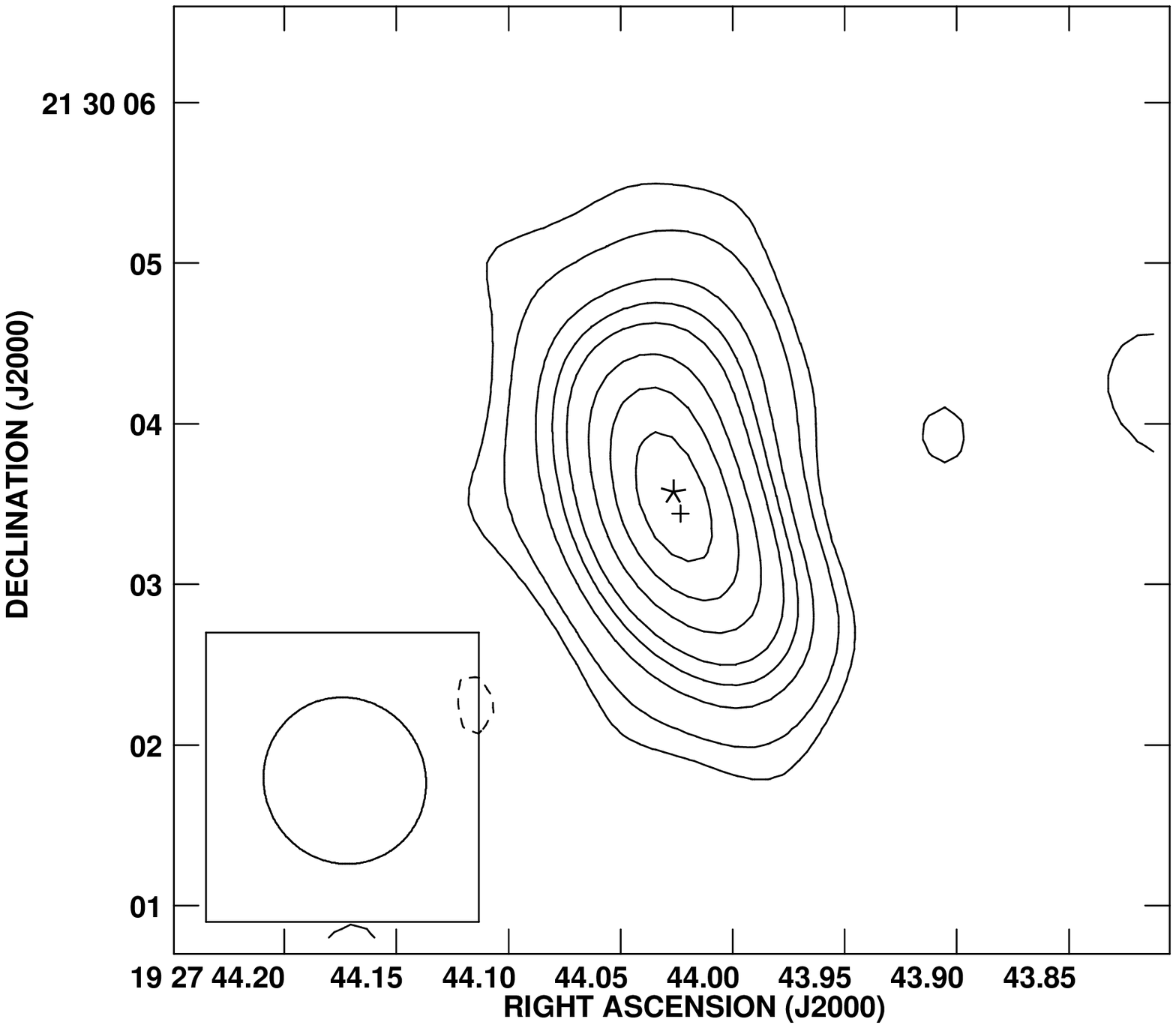}
\caption{VLA 18~cm continuum contour image of K~3-35 (beam size is
$\sim$ 1$^{\prime\prime}$, shown in the bottom left corner of the image). 
The contours are $-$3, 3, 5, 7, 9, 12, 
15, 20, and 30 times 0.3 mJy~beam$^{-1}$, the $rms$ noise of the image.
The cross marks the continuum peak position at 1.3 cm at 
$\alpha$(2000)= 19$^h$ 27$^m$ 44${\rlap.}^{s}$0233,
$\delta$(2000)= 21$^{\circ}$ 30$^{\prime}$ 03${\rlap.}^{\prime\prime}$441,
\citep{MI01, usc08}. The star marks the
peak position of the 18 cm continuum at 
$\alpha$(2000)= 19$^h$ 27$^m$ 44${\rlap.}^{s}$026,
$\delta$(2000)= 21$^{\circ}$ 30$^{\prime}$ 03${\rlap.}^{\prime\prime}$57 (this work).
Note the NE-SW elongation of the emission that follows the  same orientation 
as the bipolar outflow reported at higher frequencies \citep{MI01}. 
\label{fig1}}
 \end{center}
\end{figure}

\begin{figure}
\begin{center}
\vspace{3cm}
\includegraphics[angle=0,scale=0.7]{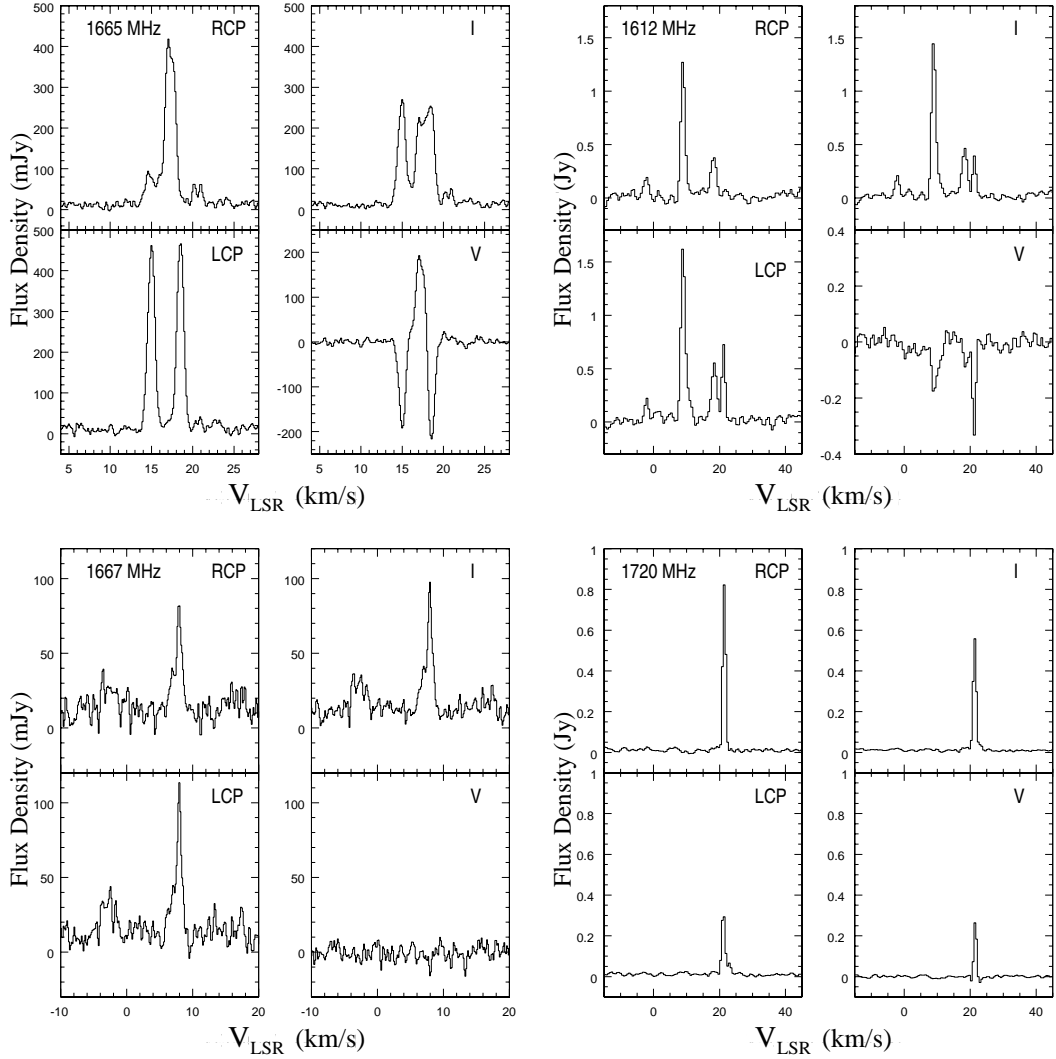}
\caption{RCP, LCP, I and V spectra toward K~3-35, for the 1665, 1667, 1612, and
1720 MHz OH transitions. The spectra were obtained by integrating the emission in
a box of 5$^{\prime\prime} \times $ 5$^{\prime\prime}$ centered on
the continuum emission peak at 18~cm. The systemic 
V$_{LSR}$ velocity of the source is $\sim$23 km~s$^{-1}$. 
Note the ``S'' profile in the Stokes parameter
V in the 1665 MHz transition at $\sim$18~km~s$^{-1}$, indicating a Zeeman pattern. 
\label{fig2}}
 \end{center}
\end{figure}

\clearpage

\begin{figure}
\begin{center}
\vspace{3cm}
\includegraphics[angle=0,scale=0.7]{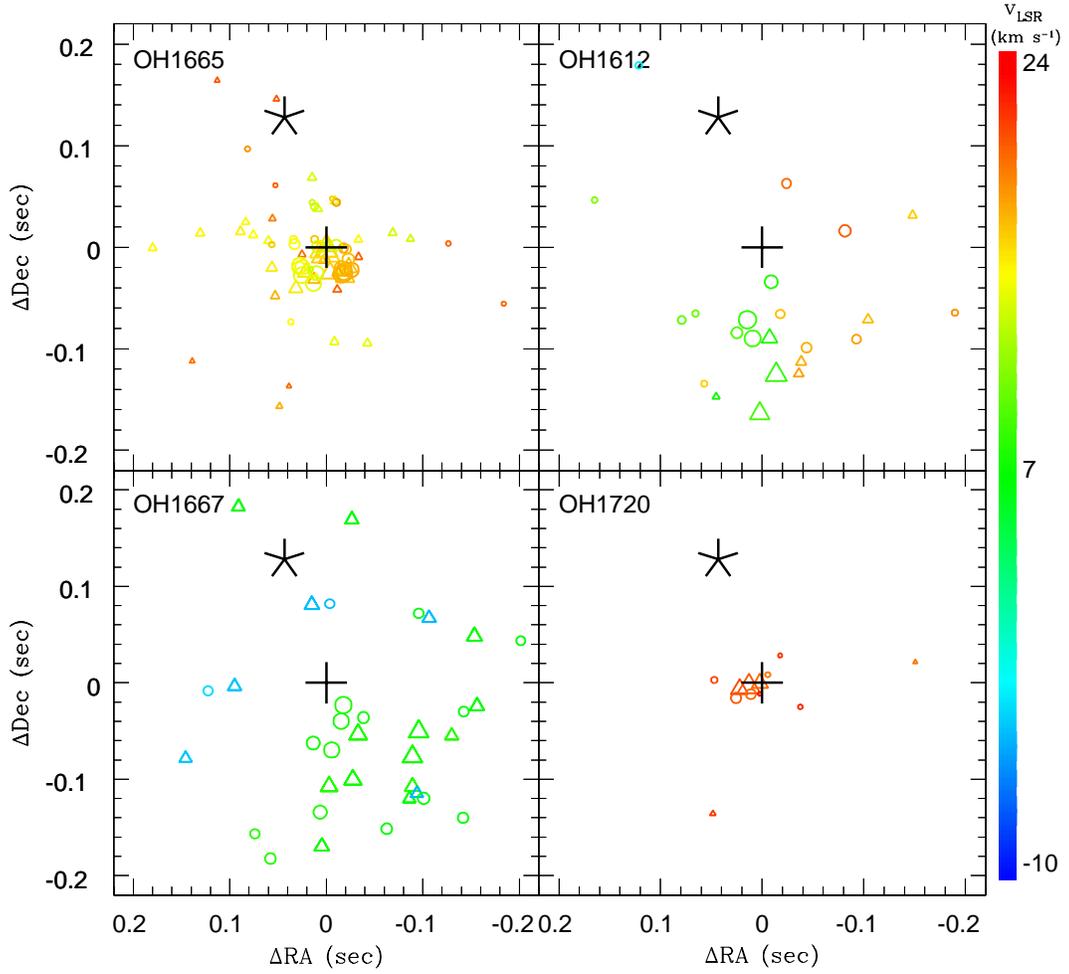}
\caption{OH maser spots for the 1665, 1667, 1612 and 1720 MHz
transitions toward the young PN K~3-35.
The circles mark the left circular polarization (LCP) and the triangles
the right circular polarization (RCP). 
Only maser spots with signal-to-noise ratio $>$5 were plotted.
The maser spots have been plotted with the same color-coded velocity
that is shown at the right side. The linear size of the symbols
is proportional to the square root of the integrated flux density.
The black cross (0,0) marks the continuum peak position at 1.3 cm 
and the black star marks the peak position of the 18 cm continuum (see \S 3.1). 
\label{fig3}}
 \end{center}
\end{figure}

\clearpage

\begin{figure}
\begin{center}
\vspace{3cm}
\includegraphics[angle=0,scale=0.8]{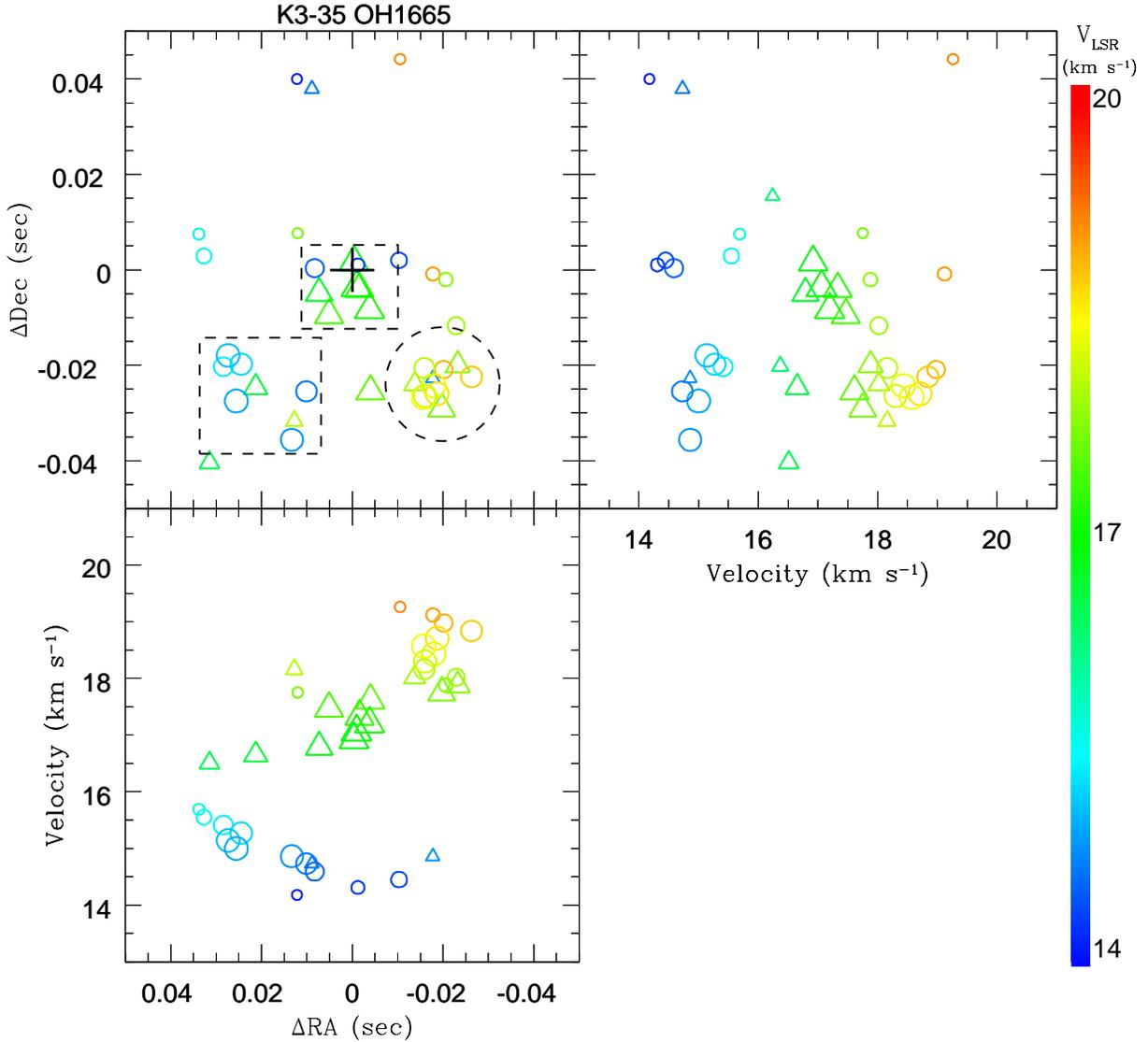}
\caption{Close-up of the position-velocity diagrams for the 1665 MHz
transition along the declination (top-right panel) and  
right ascension (bottom-left panel) of K~3-35.
Only maser spots above 15$\sigma$ were plotted, in order to note the
three different groups that have been marked with two black dash-line squares 
and one black dash-line circle.
The OH maser spots have been plot with the same color-coded velocity.
The color circles mark the left circular polarization (LCP) and the triangles
the right circular polarization (RCP) spots. The black dash-line circle marks 
the region where we suggest there is a Zeeman pair (see \S 3.2). The black cross 
indicates the 1.3 cm peak continuum position.
\label{fig4}}
 \end{center}
\end{figure}








\clearpage
\input{table1.tex}
\input{table2.tex}
\clearpage

\end{document}

%% file: table1.tex
\begin{deluxetable}{lrrr}
\tablewidth{0pt}
\tablecaption{Calibrator Flux Densities \label{tbl-1}}
\tablehead{
\colhead{Frequency}&
\multicolumn{3}{c}{Flux Density (Jy)}
\\
\cline{2-4}
\colhead{ }&
\colhead{Amplitude}&
\colhead{Phase}&
\colhead{Bandpass}
\\
\colhead{(MHz) }&
\colhead{1331+305$^a$}&
\colhead{1925+211$^b$}&
\colhead{0319+415$^b$}
}
\startdata
1665& 13.631 & 1.094$\pm$0.005 & 23.1$\pm$0.1 \\
1667& 13.623 & 1.087$\pm$0.005 & 23.4$\pm$0.1 \\
1612& 13.852 & 1.216$\pm$0.027 & \nodata $^c$ \\
1720& 13.412 & 1.069$\pm$0.007 & 23.4$\pm$0.3 \\
\enddata
\tablenotetext{a}{Adopted flux density.}
\tablenotetext{b}{Boostrapped flux density for the phase and bandpass calibrators,
respectively.}
\tablenotetext{c}{The bandpass calibrator was not observed at 1612 MHz.}
\end{deluxetable}

%% file: table2.tex
\begin{deluxetable}{ccccccrccccc}
\tabletypesize{\tiny}
\tablecolumns{12} 
\tablecaption{OH maser spectral features observed toward K~3-35$^a$ \label{tbl-2}}
\tablewidth{0pt}
\setlength{\tabcolsep}{0.10in}
\tablehead{
\multicolumn{1}{c}{}&
\multicolumn{5}{c}{RCP}&
\multicolumn{5}{c}{LCP}&
\multicolumn{1}{c}{}\\
\multicolumn{1}{c}{}&
\multicolumn{5}{c}{\hrulefill}&
\multicolumn{5}{c}{\hrulefill}&
\multicolumn{1}{c}{}\\
\colhead{Transition}& \colhead{V$_{LSR}^b$} & \colhead{S$_{\nu}$} & \colhead{$\alpha$(J2000)$^c$} & \colhead{$\delta$(J2000)$^c$} & \colhead{$\Delta^d$} & \colhead{V$_{LSR}^b$} & \colhead{S$_{\nu}$} & \colhead{$\alpha$(J2000)$^c$} & \colhead{$\delta$(J2000)$^c$} &\colhead{$\Delta^d$}&  \colhead{m$_c^e$}\\
\colhead{MHz}&
\colhead{(km~s$^{-1}$)}&
\colhead{(mJy)}&
\colhead{(s)}& 
\colhead{($"$)}&
\colhead{($"$)}&
\colhead{(km~s$^{-1}$)}&
\colhead{(mJy)}&
\colhead{(s)}&
\colhead{($"$)}&
\colhead{($"$)}&
\colhead{($\%$)}
}
\startdata
1665 &21.0 &61.6 &44.020 &03.40 &0.04 & 21.0&42.4&44.020&03.40&0.05    & +18\\
     &20.1 &66.1 &44.020 &03.40 &0.06 & 20.1&14.6&43.991&03.60&0.10    & +64\\ 
     &18.6 &37.0 &44.020 &03.30 &0.10 & 18.6&482.0&44.0200&03.400&0.005 & -86\\
     &17.1 &422.6&44.0200&03.400&0.007& 17.1&36.5&44.027&03.60&0.15    & +84\\
     &15.0 &82.1 &44.020 &03.40 &0.04 & 15.0&474.8&44.0271&03.400&0.006& -71\\
     &14.6 &99.8 &44.027 &03.40 &0.04 & 14.6&304.7&44.020&03.40&0.01   & -51\\ 
\\
1667 &8.0  &87.8&44.020  &03.40  &0.04  & 8.0 & 114.3 & 44.020  &03.40  &0.03 & -13\\
     &$-$3.5 & 43.0&44.027  &03.60  &0.07  &$-$3.5& 33.8 & 44.020  &03.60  &0.09 &+12\\
\\
1612&21.2&$<$90$^f$&\nodata&\nodata&\nodata& 21.2&774.3&44.018  &03.46  &0.02 & -90\\
     &18.4 &383.2&44.021 &03.33 &0.06 & 18.4&578.6 &44.020  &03.34 &0.04    & -20\\ 
     &8.8 &1496.6&44.022 &03.31 &0.01 & 8.8 &1815.8&44.024  &03.37 &0.01    & -10\\
     &$-$2.0&155.8 &44.039 &03.76 &0.16 &$-$2.0&261.6 &44.032  &03.62 &0.08 & -25\\ 
\\
1720 &21.4&838.7 &44.0242 &03.437 &0.003 & 21.4&306.5 &44.0251  &03.425 &0.007 & +47\\
\enddata
\tablenotetext{a}{A maser feature is defined as all the
spectral channels in which maser emission is detected, indicated by the properties of
its peak channel.}
\tablenotetext{b}{Central velocity of the channel.}
\tablenotetext{c}{Position obtained by a Gaussian fitting of the emission in each channel ($\alpha$(2000)= 19$^h$ 27$^m$ \nodata, $\delta$(2000)= 21$^{\circ}$ 30$^{\prime}$ \nodata).}
\tablenotetext{d}{Relative positional error}
\tablenotetext{e}{{\it m$_c$} Percentage of circular polarization (100$\times$V/I)}
\tablenotetext{f}{3$\sigma$ upper limit}
\end{deluxetable}